\documentclass[prc,showpacs,floatfix]{revtex4}
\usepackage{bm}
\usepackage{dcolumn}
\usepackage{graphicx}
\begin{document}
\preprint{MKPH-T-05-10}
\title{
\hfill{\small {\bf MKPH-T-05-10}}\\                                           
{\bf Signature of the $\eta NN$ configurations in coherent $\pi^0$
photoproduction on the deuteron}}
\author{A. Fix}
\affiliation{
Institut f\"ur Kernphysik,
Johannes Gutenberg-Universit\"at Mainz, D-55099 Mainz, Germany}
\date{\today}
\begin{abstract} 
The photoproduction of a neutral pion on the deuteron is considered in
the energy region around the $\eta$ threshold, where a bump-like
structure was observed at very backward pion angles. Different
dynamical aspects which may be responsible for this phenomenon are
analysed within a theoretical frame which includes intermediate $\eta
NN$ configurations. The results show in particular, that a three-body
treatment of the $\eta NN$ interaction is of special importance. 
\end{abstract}
 
\maketitle 
 

\section{Introduction}\label{intr}

In various scattering and production reactions of pions on the deuteron, 
the cross sections exhibit a bump at backward
angles for energies around the $\eta$ production threshold. In early
discussions of the underlying dynamical mechanisms the main emphasis
was put on the interpretation of these anomalies in terms of dibaryon
resonances~\cite{Akem}.  

Later work, however, was mainly focused on a more natural explanation
by relating this structure to the threshold effect caused by the
opening of the $\eta$ production channel. In particular, in a recent
experiment on coherent $\pi^0$ photoproduction such a bump was
observed in the cross section in the backward direction around a lab
photon energy $E_\gamma=700$ MeV~\cite{Yord}. 
This nontrivial energy dependence was explained
in~\cite{Kudr}, in which the authors conclude that the mechanism,
where first an $\eta$ meson is produced on one nucleon which then
interacts with the second nucleon followed by $\pi^0$ production, is
responsible for the enhancement around the $\eta$
threshold. In particular, according to the conclusion of~\cite{Kudr}, 
the appearance of the peak is nothing else
than the signature of the $S_{11}(1535)$ resonance whose excitation is
believed to dominate $\eta$ photoproduction $\gamma N\to\eta N$. 

The purpose of the present paper is to perform a more detailed and
extensive investigation of this phenomenon. The main question we
address here is the same as in~\cite{Kudr}, namely what is the
dynamics behind this enhancement. Although it might generally be clear
that the observed structure is caused by the appearance of an $\eta$
meson in one way or another, the question about the underlying
mechanism seems to be non-trivial. To be more specific, we would
firstly like to note three main points which will be discussed
separately.  

(i) The wide peak in the $\gamma d\to\pi^0 d$ cross section can be a
direct consequence of the cusp-like structure in the elementary
amplitude $\gamma N\to\pi^0 N$ in the $S_{11}$ channel near the $\eta$
threshold. The case in point is the strong coupling between the $\pi
N$ and $\eta N$ states in the region of the $S_{11}(1535)$ resonance
resulting in a very pronounced cusp in the electric dipole amplitude
$E_{0+}$ for pion photoproduction at $E_\gamma\approx$ 710 MeV. The
latter was also observed in the energy independent multipole analyses
(see, e.g., energy independent solution in Ref.~\cite{Arndt}). Turning
now to the process on the nucleon, which is bound in the deuteron, the
elementary amplitude is expected to undergo an energy shift and a
broadening of its structure due to the Fermi motion. Of crucial
importance is the question to which extent this modification could
affect the cusp in the elementary $E_{0+}$ multipole and how prominent
could be the $S_{11}(1535)$ resonance in the reaction on the deuteron.  

(ii) The bump-like structure could also arise from an additional
mechanism, appearing when the elementary photoproduction process is
embedded into the deuteron. Namely, one could expect that the
anomalies are, at least partially, caused by the three-particle
unitary cut in the amplitude $\gamma d\to\pi^0d$ starting at the
energy of the $\eta$ threshold. This cut arises because of the
possibility to exchange a physical $\eta$ meson between the 
$S_{11}$ resonance excited on one of the nucleons 
and the second nucleon. The exchange mechanism 
is characterized by the pole which turns into the
cut after the loop integration. Thus, the opening of a new physical
channel leads to an additional contribution in the imaginary part of
the amplitude reflecting a new inelasticity. Such a picture is typical
for coupled channels and, if the corresponding dynamical equations are
exactly solved, results in the three-body unitary relation. Because 
here we take into account only the leading term in the multiple
scattering series, the whole amplitude does not fulfill the unitary
relation. However, the $\eta NN$ three-body cut appears already at
this level, and the question is, how does the amplitudes behave at
the branch point. 

(iii) A sizable attraction in the $\eta NN$ system can lead to a
strong correlation between all three particles. It is worth noting
that in some work the $\eta NN$ interaction is predicted to generate
even a bound state in the quasi-deuteron configuration
$(J^\pi;T)=(1^-;0)$. Results provided by more sophisticated
models~\cite{Wycech,FiArNP} show, however, that the fundamental $\eta
N$ interaction is likely to be too weak for yielding binding of the
$\eta NN$ system, so that only a virtual (antibound) state can be
generated. In the context of the present discussion two points are
relevant: (a) Although the pole `recedes' to the nonphysical region,
it remains quite close to the zero energy. As a result, the
virtual state can strongly influence physical processes involving an
$\eta$ meson. Indeed, a variety of theoretical calculations and
experimental analyses exhibit a strong rise of the $\eta$ production
cross section just above threshold. This collective effect, in which
all three particles participate, is naturally explained within a
three-body model~\cite{FiArNP}. In this sense, the origin of the pole
in the amplitude is not of crucial significance, and the existence of
an $\eta NN$ bound state is not the necessary condition for an
anomalous behavior of the $\eta$ production cross section. Important
is only how far is the pole from the threshold energy. (b)
Perturbative models, like the first order rescattering approximation
noted in (ii), where only the leading order terms are kept in the
multiple scattering series, are unable to reproduce the real dynamics
of the $\eta NN$ system in the low energy region. The most simple
explanation is that the corresponding Neumann series converges very
slowly near the pole position~\cite{Schmid}, so that the leading terms
turn out to be a bad approximation to the whole series. From the last
considerations we conclude that any realistic study of the role of
$\eta$ mesons in the reaction $\gamma d\to\pi^0 d$ should be based on
a three-body approach to the $\eta NN$ system. 

Resuming now the qualitative discussion, we would like to note that
all mentioned factors can come into play to form the observed
characteristic bump in coherent $\pi^0$ photoproduction close to
the $\eta$ threshold. Important is the quantitative relation between
the different mechanisms, which is the main object of the present
paper. In the following we consider all three points separately
with special emphasis with respect to their contribution to the
resulting cross section. A brief description of the formal ingredients
in Sect.~\ref{form} is followed by the discussion of our results for
the differential cross section of $\gamma d\to\pi^0 d$ in
Sect.~\ref{res}. An appendix contains the listing of various formulas
used for the calculation of the reaction amplitude in the impulse
approximation. 


\section{The Formalism}\label{form}

\begin{figure}[h]
\includegraphics[width=14cm]{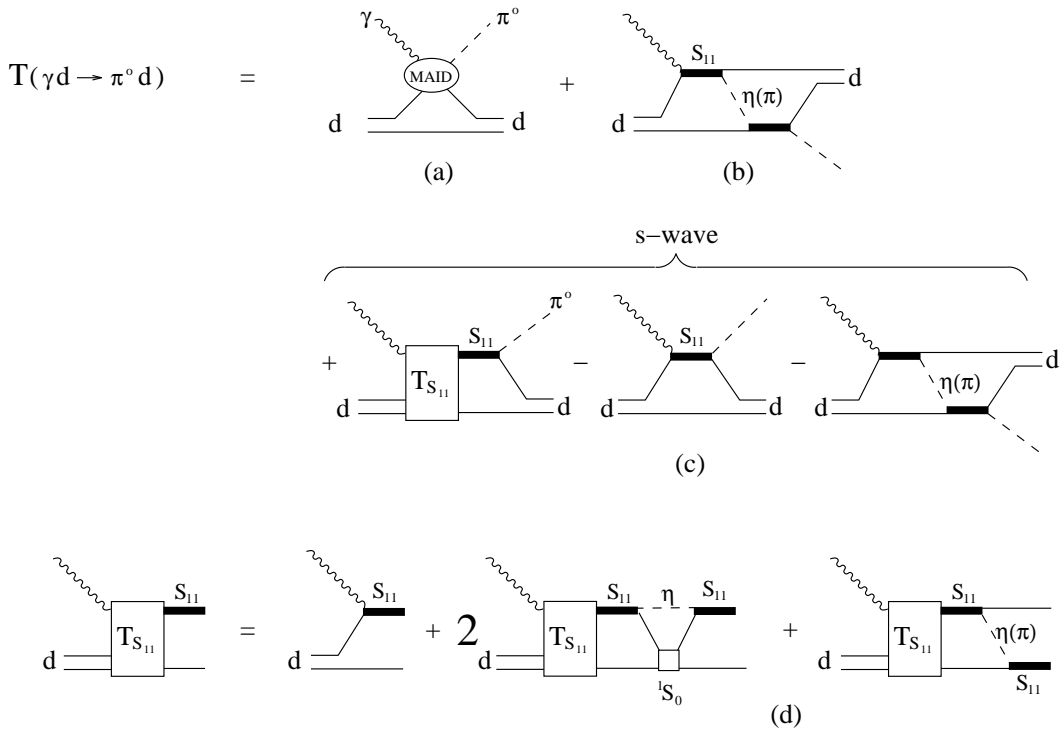}
\caption{Diagrams for the reaction $\gamma d\to\pi^0 d$
included in the present work: (a) impulse approximation; (b) first order 
rescattering contribution; (c) additional contribution from three-body 
dynamics only in the $s$-wave ($J^\pi,T=0^-,1$); (d) equation for the 
amplitude $T_{S_{11}}$ 
of photoproduction of the $S_{11}$ resonance on the deuteron.}
\label{fig1}
\end{figure}

We start the brief description of the basic formal ingredients by
presenting in Fig.~\ref{fig1} the diagrams which we consider in the
theoretical analysis. The three combinations (a), (a)+(b), and
(a)+(b)+(c) present three different, successively improved levels of
approximation to the reaction amplitude, corresponding to the three
points (i) through (iii) discussed above. We will refer to them as
impulse approximation (IA), first order rescattering approximation,
and three-body calculation, respectively. 

The starting point of our formalism is the impulse approximation which
is completely determined by the elementary amplitude for $\gamma
N\to\pi^0 N$ and the deuteron wave function. As elementary amplitude
we take the MAID analysis~\cite{MAID} and for the deuteron wave
function the parametrization of the Bonn-potential model
(OBEPQ-version)~\cite{Bonn} with inclusion of the tensor
component. The latter is certainly very important in the region of
large momentum transfers. The general expression for the amplitude is
given in the Appendix. 

For the additional contributions of the diagrams (b) and (c), we adopt
the following simplifications. Firstly, in the pion exchange
contribution to the second diagram only the $S_{11}$ resonance was
taken into account neglecting the contribution of other
resonances. Although this neglect leads to an underestimation of the
pion rescattering effect it should not strongly affect the quality of
our results because of the following reasons: (i) since rescattering
of intermediate pions is not related to the threshold effects it does
not contribute into formation of the peak structure which we discuss
here. What we can expect is only a smooth change of the cross section
at backward angles; (ii) the role of pion exchange seems to be not
very essential. In Ref.~\cite{Garc} where this mechanism was
calculated more precisely the corresponding effect in the second
resonance region is less than 20~\%. 

Secondly, as is indicated in Fig.~\ref{fig1}, the three-body problem
for the $\eta NN$ system was solved only for the $s$-wave state
$^1S_0$ ($J^\pi;T$ = $0^-;1$). As is shown in Ref.~\cite{FiArNP} it is
the state of lowest orbital momentum which is mostly distorted by the
multiple scatterings between particles. Other states with higher
orbital momentum can be included perturbatively within the
rescattering approximation (diagrams (a)+(b)). In order to avoid
double counting we remove from the three-body amplitude those diagrams
which possess the same topology as the ones already included in (a) and
(b) shown in the same figure. 

For the elementary amplitudes appearing in other diagrams (b) and (c) 
of Fig.~\ref{fig1}, we assume that the photoproduction of an $\eta$
meson as well as its interaction with nucleons proceeds exclusively
via the excitation of the $S_{11}(1535)$ resonance. As mentioned
above, the same ansatz was adopted for pions. According to this
assumption the $t$ matrix of meson-nucleon scattering is given by the
conventional isobar model 
\begin{equation}\label{tmat}
t_{\alpha N\to\beta N}(\vec{p},\vec{p}\,';W)=
\frac{g_\alpha(\vec{p}\,)g_\beta(\vec{p}\,')}{W-M_0-\Sigma_\eta-\Sigma_\pi
-\Sigma_{\pi\pi}}\,,\quad \alpha,\beta \in\{\pi,\eta\}\,, 
\end{equation}
where $W$ denotes the invariant energy and $\Sigma_\alpha$ the various 
self energy contributions from the $\alpha N$ channels with 
$\alpha\in\{\pi,\eta,\pi\pi\}$. The $t$ matrix is determined by
the bare resonance mass $M_0$ and the parameters of the vertex
functions $g_\alpha$, for which we take a simple Hulth\'en form
\begin{equation}
g_\alpha(\vec{p}\,)=g_\alpha\left(1+\frac{p^2}{\Lambda_\alpha^2}\right)^{-1}\,.
\end{equation}
The contributions to the self energy from the various channels are
expressed in terms of $g_\alpha(\vec{p}\,)$ as 
\begin{equation}
\Sigma_\alpha(W)=\frac{1}{2\pi^2}\int\limits_0^\infty\frac{g_\alpha(q)^2}
{W-E_N(q)-\omega_\alpha(q)+i\varepsilon}\frac{q^2dq}{2\omega_\alpha}
\end{equation}
for $\alpha\in\{\pi,\eta\}$. Since the double pion channel $\pi\pi N$
is not explicitly included in the calculation, primarily because of
its rather weak coupling to the $S_{11}(1535)$, we parametrize,
following~\cite{BeTa}, the corresponding self energy in a simplified
manner as a pure imaginary contribution proportional to the
three-particle phase space 
\begin{equation}
\Sigma_{\pi\pi}=-\frac{i}{2}\gamma_{\pi\pi}\frac{W-M_N-2m_\pi}{m_\pi}\,.
\end{equation}

For the photoproduction amplitude we take the same ansatz as in
(\ref{tmat}) where one hadronic vertex function is replaced by the
electromagnetic vertex $g_{\gamma N}$ for $\gamma N\to S_{11}$ which
depends only on the invariant energy $W$ and is parametrized in the
form 
\begin{eqnarray}\label{em1}
g_{\gamma p}(W)&=&\left\{
\begin{array}{ll}\displaystyle 
\frac{e}{\sqrt{4\pi}}
\sum\limits_{n=0}^4 a_n\left(\frac{q_\pi}{m_\pi}\right)^n, &
\quad \mbox{for} \,\,W\ge M_N+m_\pi\,,\\
g_{\gamma p}(M_N+m_\pi)\,, & \quad \mbox{else}\,,
\end{array}
\right.\\
&&\nonumber\\
g_{\gamma n}(W)&=&-0.82\ g_{\gamma p}(W)\,.
\nonumber 
\end{eqnarray}
with $q_\pi=\sqrt{(W^2-(M_N+m_\pi)^2)(W^2-(M_N-m_\pi)^2)}/2W$.
The isospin dependence of the $S_{11}(1535)$
photo excitation amplitude was taken according to the 
relation~\cite{Hej99}
\begin{equation}\label{em2}
\frac{\sigma(\gamma p\to\eta p)}
{\sigma(\gamma n\to\eta n)}\approx 0.67\,.
\end{equation}

\begin{table}
\renewcommand{\arraystretch}{2.0}
\caption{
Parameters of the $\eta N$ scattering matrix in Eqs.~(\protect\ref{tmat}) 
through (\protect\ref{em1}). The values of $\Lambda_\alpha$, $\gamma
_{\pi\pi}$, and $M_0$ are in MeV.} 
\begin{ruledtabular}
\begin{tabular}{ccccccccccc}
$g_\eta$ & $\Lambda_\eta$ & $g_\pi$ & 
$\Lambda_\pi$ &$\gamma_{\pi\pi}$ & $M_0$ 
& $a_0$ & $a_1$ & $a_2$ & $a_3$ & $a_4$ 
\\ \colrule
\ 2.00\ &\ 694.6\ &\ 2.51\ &\ 404.5\  &\ 4.3\ &\ 1598\ &\ 
$5.502\cdot 10^{-1}$ & 
$-1.923\cdot 10^{-2}$ & $1.018\cdot 10^{-1}$ & $2.255\cdot 10^{-3}$ & 
$-7.042\cdot 10^{-3}$\\
\end{tabular}
\end{ruledtabular}
\label{tab1}
\end{table}

The parameters, appearing in the expressions (\ref{tmat}) through
(\ref{em1}), are listed in Table~\ref{tab1}. They are chosen in such a
way that, on the one hand, the reactions $\gamma N\to\alpha N$ and 
$\pi^-p\to\eta n$ are well reproduced in the
$S_{11}$ channel as presented in our previous
work~\cite{FiArDo,FiAr3N}. On the other hand, the chosen parameter set
predicts for the $\eta N$ scattering length the value $a_{\eta
N}=(0.5+i0.3)$~fm which we consider as an approximate average of the
various values provided by the $\eta N$ analyses. 

The method which we use to solve the three-body problem for the $\eta
NN$ interaction is described in Ref.~\cite{FiArNP} and we refer the
reader to this paper for the details. Here we only would like to
mention that the key point of the method is the separable
representation of the driving two-body interaction in the $\pi N$,
$\eta N$, and $NN$ subsystems. The corresponding $t$ matrix for
meson-nucleon multiple channel scattering is given by the isobar
formula (\ref{tmat}). For the $NN$ sector, we use the separable
representation of the Bonn potential as given in Ref.~\cite{Haid} for
the $^1S_0$ and $^3S_1$ configurations. 

As is well known, the separable ansatz makes it
possible to reformulate the three-body problem in terms of two-body
scattering between quasiparticles. As a consequence, the solution of
the problem is given by an amplitude $T_{S_{11}}$
of the effective transition
$\gamma d\to NS_{11}$ as presented in Fig.~\ref{fig1}(d). 
The needed physical amplitude $\gamma d \to\pi^0d$ is then obtained 
through an additional loop integration 
(the first diagram in Fig.~\ref{fig1}(c)).


\section{Results and discussion}\label{res}

\begin{figure}[h]
\includegraphics[width=14cm]{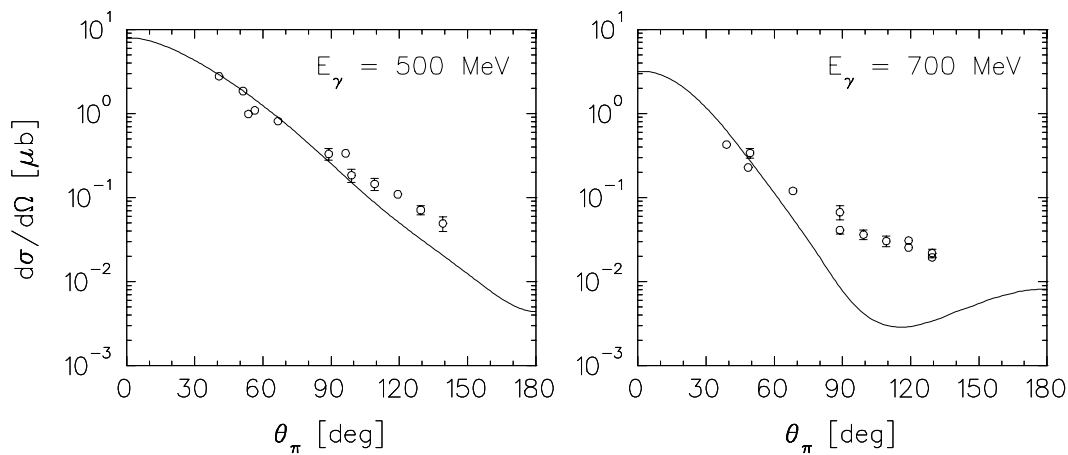}
\caption{Angular distribution for $\gamma d\to\pi^0d$ at two 
photon energies. The curves are the results of the impulse approximation (IA). 
The data are taken from the compilation in~\protect\cite{NuKa}.}
\label{fig2}
\end{figure}

In order to demonstrate the quality of the impulse approximation, we
present in Fig.~\ref{fig2} the resulting pion angular distribution,
and compare them  with the available data from the compilation
of~\cite{NuKa}. Although the agreement is quite satisfactory for the
forward angles, the data
are significantly underestimated in the backward region. We will 
return to this point at the
end of the discussion. A comparison with the IA calculations of
Ref.~\cite{Garc} also exhibits differences which are, however, not very
dramatic. For instance, our cross section at $E_\gamma$ = 
700 MeV shows quite a monotonic behavior and does not possess a deep minimum at
very forward angles as obtained in Ref.~\cite{Garc}. It is intuitively 
clear that
as long as small angles are considered where the two-nucleon effects are
minimal (except for pion two-body absorption) the magnitude of the
cross section should be mostly determined by the elementary amplitude. 
Therefore, the difference between the two theoretical results has to
be ascribed primarily 
to the differences in the corresponding elementary operators, especially in the
spin-flip part (see Eq.~(\ref{KL})), which accounts for the
dominant fraction of the forward cross section on the deuteron. 

\begin{figure}
\includegraphics[width=14cm]{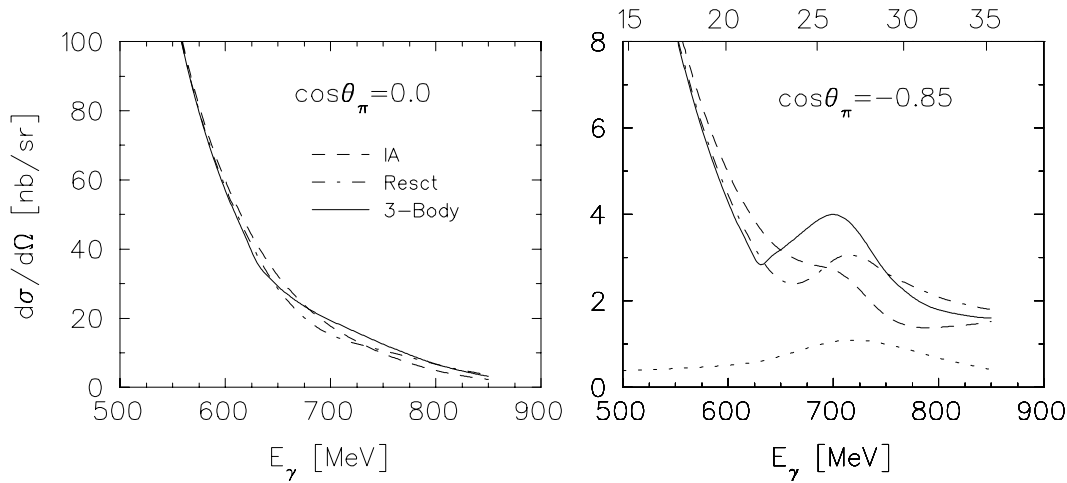}
\caption{Differential cross section for $\gamma d\to\pi^0d$ as
function of incident photon energy for two pion emission angles in the
$\gamma d$ c.m.\ system. The curves show the results of the impulse 
approximation
(dashed), first order rescattering (dash-dotted) and three-body calculation
(solid). The dotted curve shows the contribution of the $\eta$
rescattering term alone (diagram (b) in Fig.~\protect\ref{fig1}) where 
the resonance propagators were substituted by constants (see text). The 
corresponding values of the four-momentum transfer squared in fm$^{-2}$ 
are shown on the top abscissa.}
\label{fig4}
\end{figure}

Now we turn to our main results on the additional interaction effects
as presented in Fig.~\ref{fig4}. The curves show the predictions
according to the different approximations discussed in the
introduction. As one might expect, at forward angles the cross section
shows very little influence of these effects. With
increasing momentum transfer, corresponding to increasing pion
emission angles, small internuclear distances come into play and thus
corrections to the simple IA calculation from the two-nucleon
mechanisms become more and more important. Furthermore, a bump in the
energy dependence around $E_\gamma$ = 650 MeV is clearly visible at
very backward angles. Among other things, this fact can be considered
as strong evidence that primarily the two-nucleon mechanisms are
responsible for this phenomenon. 

This obvious statement does not, however, diminish the role of the
single nucleon response. As one readily sees in the right panel of
Fig.~\ref{fig4}, some nontrivial structure, a shoulder, appears in the
cross section near the $\eta$ threshold already in the impulse
approximation where the second nucleon is not actively involved. A
more detailed analysis reveals, however, that the enhancement
is not caused by the presence of the $S_{11}(1535)$ resonance. 
Rather it is a combined effect of different
terms in the MAID amplitude which we use here. 

On the other hand, we would like to note that the $S_{11}(1535)$
resonance itself does in principle produce a slight shoulder close to
the $\eta$ threshold, which, as already mentioned in the introduction,
is a signature of the cusp in the $E_{0+}$ multipole
smeared out by the Fermi motion in the deuteron. However, this
resonance contributes little to the coherent reaction on the
deuteron, so that this cusp-like structure turns out to be invisible
in the cross section. Hence, the slight enhancement observed in the
otherwise monotonic behavior of the IA cross section is not related to
the $\eta NN$ dynamics and should be ascribed to properties of the
elementary pion production amplitude in this region. 

Now, when the first-order rescattering of the produced $\pi$ and
$\eta$ mesons is included, a peak structure clearly evolves as
exhibited by the dash-dotted curve in Fig.~\ref{fig4}. As was
emphasized in~\cite{Kudr}, the main origin of this effect is the
presence of the $S_{11}(1535)$ resonance in the diagram with $\eta$
rescattering. Our results in general confirm this statement. It is,
however, interesting that the new three-body unitary cut in the
amplitude connected with the $\eta NN$ channel also make a slight
contribution to the formation of the bump. Indeed, if the strong
energy dependence of the amplitudes $\gamma N\to\eta N$ and $\eta
N\to\pi^0 N$ (see diagram (b) in Fig.~\ref{fig1}) near the $\eta$
threshold is `neutralized' by keeping the $S_{11}$ propagator
constant, the rescattering term alone exhibits slight enhancement 
(dotted line in Fig.~\ref{fig4}). 

Turning now back to the discussion of three-body effects in $\pi^0$
photoproduction, the three-body calculation clearly predicts a much
more prominent peak structure accompanied by a slight shift to lower
energies. Thus the difference between the solid and the dash-dotted
curves demonstrates convincingly the importance of the higher order
terms in the multiple scattering series for the intermediate $\eta NN$
interaction. It is worth noting that in the `ideal case' the
measurement of the peak might be an indicator of the dynamical
properties of the $\eta NN$ system. In particular, if a bound $\eta
NN$ state would exist, it would appear in the $\pi NN$ channel as a
pure $s$-wave three-body resonance. The mass difference
$2M_N+m_\eta-M_R$, where $M_R$ is the mass of this hypothetic
resonance, would then give the binding energy. If on the other hand
the $\eta NN$ scattering amplitude possesses only a virtual pole, as
is predicted by our model, the resonance peak should fall directly on
the $\eta$ threshold. Nevertheless, it is obvious that the extraction
of such information would require a quite thorough partial wave
analysis (we need to consider only the $s$-wave contribution) and,
moreover, a very precise energy resolution and a small statistical
error, which could then allow one to fix the resonance position. A
similar scheme was already realized in the experiments aimed at a
search for $\eta$ mesic nuclei with $^3$He~\cite{Pfeif} and heavier
targets~\cite{Sok}. But although the quality of the data permits one
to make quite definite conclusion about the peak position, there are
still problems with the interpretation of the measured angular
distribution and the energy dependence of the corresponding cross
section within existing theoretical models~\cite{Pfeif}. 

\begin{figure}
\includegraphics[width=14cm]{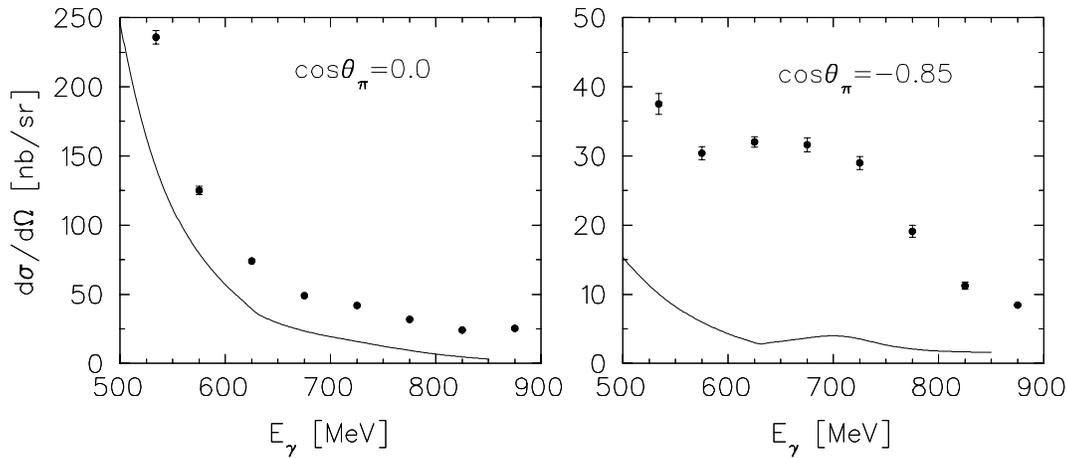}
\caption{Comparison of our three-body results (solid line in 
Fig.~\protect\ref{fig4}) with CLAS preliminary data~\cite{Yord}.}
\label{fig5}
\end{figure}
A comparison of our results in Fig.~\ref{fig5} with preliminary data
from Ref.~\cite{Yord} shows that the theory underestimates
the observed cross section at backward angles by about one order of
magnitude. The same discrepancy is noted for the results of
Ref.~\cite{Kudr} where the cross section is also far below the data in
the same region. Yet, the authors of~\cite{Kudr} use an oversimplified
operator for $\gamma N\to\pi N$ and it is therefore difficult to
identify the prime reason of this drawback. Furthermore, similar
underestimation may be exhibited in the work of~\cite{Garc}, although
the deviation is not as significant as in our case. 

The noted discrepancy might not be very surprising after all. 
We can expect that as long as the pions
are produced on the individual nucleons and as long as the angular
dependence of the elementary amplitude $\gamma N\to\pi^0 N$ is not
varying strongly, the form of the differential cross section is mainly
governed by the deuteron form factor. In order to locate the
appropriate portion of the form
factor which enters in the kinematical conditions, we present on
the top x-axis in Fig.~\ref{fig4} the corresponding four
momentum transfer squared. As one can see, its characteristic values
range between 15 and 35 fm$^{-2}$, where the form factor as seen in
electron scattering exhibits a sizeable sensitivity to higher order
mechanisms like $\pi$ and $\rho$ MEC's (see, e.g., Ref.~\cite{ArRi}). 

Therefore it is very probable that additional two-nucleon mechanisms,
not included in the present calculation, will make sizeable
contributions to the backward cross section. For example, meson
exchange currents were found to be quite significant for $\pi^+$ 
photoproduction on $^3$He in Ref.~\cite{OsKa}. In this case, the effect 
is associated with double meson photoproduction followed by absorption 
of one of the mesons on the spectator nucleon. However, if such 
mechanisms are really quite important, their inclusion would tend to 
diminish the effect discussed above, i.e.\ the relative size of the 
peak as its manifestation. In this case, we will have to face again 
the same question about the origin of the structure observed in the 
$\pi^0$ photoproduction.


\section{Conclusion}

We have discussed various aspects of the influence of $\eta$ degrees of
freedom on coherent $\pi^0$ photoproduction on the deuteron. Two
interrelated features determine the significance of the $\eta$ meson in this
reaction. Firstly, the strong coupling between $\pi N$ and $\eta N$
states in the energy region of the $S_{11}(1535)$ resonance 
leads to a significant admixture of the
$\eta NN$ configuration to the $NS_{11}$ intermediate states. What is more
important, in contrast to pions, slow $\eta$ mesons interact strongly with
nearby nucleons. As predicted in a variety of investigations, 
this dynamical feature leads to highly correlated $\eta NN$ states, 
which should manifest themselves in the $s$-wave part of the $\pi^0$
photoproduction amplitude. 

Both factors lead to the appearance of a pronounced bump in the energy
dependence of the backward differential cross section around the 
$\eta$ production threshold. This effect was analysed in Ref.~\cite{Kudr} 
on the basis of a theoretical model where in addition to the mere impulse 
approximation $\pi$ and $\eta$ rescatterings were included. In contrast to the 
conclusion of~\cite{Kudr}, we find that already in the impulse approximation
 a shoulder appears in the cross section around $E_\gamma$ = 700 MeV. However,
as is explained in the present work, this effect does not have a deep physical
significance, and is likely to reflect the special structure of the pion 
production amplitude used here. 

The inclusion of first-order rescattering and, finally, all terms in the 
multiple scattering series within a three-body model shifts the peak
position and make it significantly more pronounced. In particular, it 
was shown that a three-body treatment of the $\eta NN$ interaction is of 
special importance for the understanding of the reaction dynamics. 

In general, according to our results, the physics behind the bump structure 
in the cross section for $\gamma d\to\pi^0 d$ may be much more complicated
than was presented in Ref.~\cite{Kudr}. Among other things, we are very 
sceptical about the possibility to extract a model independent information 
on the fundamental $\eta N$ interaction from the $(\gamma,\pi^0)$ reactions. 

The present calculation as well as those presented in Ref.~\cite{Kudr} 
can be considered as a natural explanation of the experimental results
reported in~\cite{Yord}. However, the status of these findings 
is quite unclear, because the theory strongly underestimates the data in the
relevant angular region, making any quantitative analysis of the observed 
cross section impossible. Further theoretical investigations in this field are
certainly needed. 

\section*{Acknowledgements}
The author would like to thank Hartmuth Arenh\"ovel for useful discussions and 
a critical reading of the manuscript. This work was supported by the Deutsche 
Forschungsgemeinschaft (SFB 443). 


\appendix
\renewcommand{\theequation}{A\arabic{equation}}
\setcounter{equation}{0}
\section{Impulse approximation for the amplitude 
$\gamma d\to\pi^0 d$}\label{app}
Here we give a brief outline of the impulse approximation of the 
$T$-matrix of coherent pion photoproduction on the deuteron in the c.m.\ frame
\begin{equation}
\gamma(\omega_\gamma,\vec{k},\lambda)+d(E_d,-\vec{k}\,)\to
\pi(\omega_\pi,\vec{q}\,)+d(E_d',-\vec{q}\,)\,,
\end{equation}
where energy and momenta of the participating particles are given in the
parentheses, and $\lambda$ stands for the circular photon polarization index. 
In the impulse approximation, the amplitude $T_{mm'\lambda}$ 
for the transition between the target states with spin projections 
$m$ and $m'$ on the $z$-axis, chosen along the photon momentum, 
\begin{equation}\label{TMM}
T_{mm'\lambda}=2\int\frac{d^3p}{(2\pi)^3}\ \phi_{m'}^\dagger
(\vec{p}\,')\,t^\lambda_{\pi\gamma}(\vec{k},\vec{p}_i,\vec{q},
\vec{p}_f)\phi_m(\vec{p}\,)\,.
\end{equation}
with $t^\lambda_{\pi\gamma}$ standing for the corresponding elementary
amplitude $\gamma N\to\pi^0N$. Furthermore, the vectors $\vec{p}_{i}$ 
and $\vec{p}_{f}$ denote initial and final momenta of the active 
nucleon in the deuteron, for which we have $\vec{p}_i=\vec{p}-\vec{k}/2$ 
and $\vec{p}_f=\vec{p}-\vec{q}+\vec{k}/2$, and 
$\vec{p}\,'=\vec{p}+(\vec{k}-\vec{q}\,)/2$ denotes the relative momentum 
in the final deuteron state. 

For the deuteron wave function we use the familiar ansatz
\begin{equation}
\phi_m(\vec{p}\,)=\sum\limits_{L=0,2}\sum\limits_{m_Lm_S}(Lm_L1m_S|1m)
u_L(p)Y_{Lm_L}(\hat{p})\chi_{m_S}\zeta_0\,,
\end{equation}
where the last two terms denote spin and isospin wave functions, respectively. 

For nuclear application it is convenient to split the amplitude
$t_{\pi\gamma}$ into spin independent and spin-flip parts  
(the index $\lambda$ is omitted for convenience in the following expressions)
\begin{equation}\label{KL}
t_{\pi\gamma}=K+i\vec{L}\cdot\vec{\sigma}\,.
\end{equation}
Then, using standard angular momentum algebra, the reaction amplitude
(\ref{TMM}) can be put into the following form 
\begin{eqnarray}
T_{mm'}&=&A\sqrt{3}\sum\limits_{\Lambda=0}^2
(-1)^{M_\Lambda}\sqrt{2\Lambda+1}(1m\Lambda-M_\Lambda|1m')
\sum\limits_{LL'=0,2}\int\frac{d^3p}{(2\pi)^3}\Big[
\left\{\begin{array}{ccc}
L & 1 & 1 \\
1 & L'& \Lambda 
\end{array}\right\} 
\{Y^{[L']}(\hat{p}')\otimes Y^{[L]}(\hat{p})\}^{[\Lambda]}_{M_\Lambda}K
\nonumber
\\
&-& (-1)^\Lambda\sqrt{6}\sum\limits_{l=0}^3\sqrt{2l+1}
\left\{\begin{array}{ccc}
\Lambda & 1 & 1 \\
1 & 1 & 1 \\
l & L & L'
\end{array}\right\} 
\{\{Y^{[L']}(\hat{p}')\otimes Y^{[L]}(\hat{p})\}^{[l]}\otimes 
L^{[1]}\}^{[\Lambda]}_{M_\Lambda}\Big]
u_{L'}(\vec{p}\,')u_L(\vec{p}\,)\,.
\end{eqnarray}

As for the isospin structure, it easy to understand that from all three
amplitudes in the isospin decomposition of the elementary operator for
pion photoproduction with Cartesian index $\alpha$=1,\,2,\,3 \cite{CGLN}
\begin{equation}
t_{\pi\gamma}=M^{(0)}\tau_\alpha+M^{(-)}\frac{1}{2}[\tau_\alpha,\tau_3]+
M^{(+)}\delta_{\alpha 3}\,,
\end{equation}
only $M^{(+)}$ can contribute to the coherent process on the deuteron.  

Using standard normalization of particle states, the 
cross section related to the same c.m.\ frame reads 
\begin{equation}
\frac{d\sigma}{d\Omega}=\frac{q}{\omega_\gamma}\,\frac{E_dE_d'}{(4\pi W)^2}\,
\frac{1}{6}\sum\limits_{mm'\lambda}|T_{mm'\lambda}|^2\,.
\end{equation}


\end{document}